


\font\twelverm=cmr10  scaled 1200   \font\twelvei=cmmi10  scaled 1200
\font\twelvesy=cmsy10 scaled 1200   \font\twelveex=cmex10 scaled 1200
\font\twelvebf=cmbx10 scaled 1200   \font\twelvesl=cmsl10 scaled 1200
\font\twelvett=cmtt10 scaled 1200   \font\twelveit=cmti10 scaled 1200
\font\twelvesc=cmcsc10 scaled 1200
\skewchar\twelvei='177   \skewchar\twelvesy='60


\def\twelvepoint{\normalbaselineskip=12.4pt plus 0.1pt minus 0.1pt
  \abovedisplayskip 12.4pt plus 3pt minus 9pt
  \belowdisplayskip 12.4pt plus 3pt minus 9pt
  \abovedisplayshortskip 0pt plus 3pt
  \belowdisplayshortskip 7.2pt plus 3pt minus 4pt
  \smallskipamount=3.6pt plus1.2pt minus1.2pt
  \medskipamount=7.2pt plus2.4pt minus2.4pt
  \bigskipamount=14.4pt plus4.8pt minus4.8pt
  \def\rm{\fam0\twelverm}          \def\it{\fam\itfam\twelveit}%
  \def\sl{\fam\slfam\twelvesl}     \def\bf{\fam\bffam\twelvebf}%
  \def\mit{\fam 1}                 \def\cal{\fam 2}%
  \def\sc{\twelvesc}               \def\tt{\twelvett}
  \def\sf{\twelvesf}
  \textfont0=\twelverm   \scriptfont0=\tenrm   \scriptscriptfont0=\sevenrm
  \textfont1=\twelvei    \scriptfont1=\teni    \scriptscriptfont1=\seveni
  \textfont2=\twelvesy   \scriptfont2=\tensy   \scriptscriptfont2=\sevensy
  \textfont3=\twelveex   \scriptfont3=\twelveex  \scriptscriptfont3=\twelveex
  \textfont\itfam=\twelveit
  \textfont\slfam=\twelvesl
  \textfont\bffam=\twelvebf \scriptfont\bffam=\tenbf
  \scriptscriptfont\bffam=\sevenbf
  \normalbaselines\rm}



\def\beginlinemode{\endmode
  \begingroup\parskip=0pt \obeylines\def\\{\par}\def\endmode{\par\endgroup}}
\def\beginparmode{\endmode
  \begingroup \def\endmode{\par\endgroup}}
\let\endmode=\par
{\obeylines\gdef\
{}}
\def\singlespace{\baselineskip=\normalbaselineskip}

\def\oneandahalfspace{\baselineskip=\normalbaselineskip
  \multiply\baselineskip by 3 \divide\baselineskip by 2}
\def\doublespace{\baselineskip=\normalbaselineskip \multiply\baselineskip by 2}

\newcount\firstpageno
\firstpageno=2
\footline={\ifnum\pageno<\firstpageno{\hfil}\else{\hfil\twelverm\folio\hfil}\fi}

\def\toppageno{\global\footline={\hfil}\global\headline
  ={\ifnum\pageno<\firstpageno{\hfil}\else{\hfil\twelverm\folio\hfil}\fi}}
\let\rawfootnote=\footnote              
\def\footnote#1#2{{\rm\singlespace\parindent=0pt\parskip=0pt
  \rawfootnote{#1}{#2\hfill\vrule height 0pt depth 6pt width 0pt}}}
\def\raggedcenter{\leftskip=4em plus 12em \rightskip=\leftskip
  \parindent=0pt \parfillskip=0pt \spaceskip=.3333em \xspaceskip=.5em
  \pretolerance=9999 \tolerance=9999
  \hyphenpenalty=9999 \exhyphenpenalty=9999 }
\def\dateline{\rightline{\ifcase\month\or
  January\or February\or March\or April\or May\or June\or
  July\or August\or September\or October\or November\or December\fi
  \space\number\year}}
\def\received{\vskip 3pt plus 0.2fill
 \centerline{\sl (Received\space\ifcase\month\or
  January\or February\or March\or April\or May\or June\or
  July\or August\or September\or October\or November\or December\fi
  \qquad, \number\year)}}


\hsize=6.5truein
\hoffset=0.0truein
\vsize=8.5truein
\voffset=0.25truein
\parskip=\medskipamount
\toppageno
\twelvepoint
\doublespace
\def\\{\cr}
\overfullrule=0pt 




\def\title#1{                   
   \null \vskip 3pt plus 0.3fill \beginlinemode
   \doublespace \raggedcenter {\bf #1} \vskip 3pt plus 0.1 fill}

\def\author                     
  {\vskip 3pt plus 0.1fill \beginlinemode \doublespace \raggedcenter}

\def\affil                      
  {\vskip 3pt \beginlinemode \doublespace \raggedcenter \it}

\def\abstract                   
  {\vskip 3pt plus 0.1fill \subhead {Abstract:}
   \beginparmode \narrower \oneandahalfspace }

\def\endtopmatter               
  {\vskip 3pt plus 0.1fill \endpage \body}

\def\body                       
  {\beginparmode}               

\def\head#1{                    
   \goodbreak \vskip 0.4truein  
  {\immediate\write16{#1} \raggedcenter {\sc #1} \par}
   \nobreak \vskip 3pt \nobreak}

\def\subhead#1{                 
  \vskip 0.25truein             
  {\raggedcenter {\it #1} \par} \nobreak \vskip 3pt \nobreak}

\def\beneathrel#1\under#2{\mathrel{\mathop{#2}\limits_{#1}}}

\def\refto#1{${\,}^{#1}$}       

\newdimen\refskip \refskip=0pt
\def\references         
  {\head{References}    
   \beginparmode \frenchspacing \parindent=0pt \leftskip=\refskip
   \parskip=0pt \everypar{\hangindent=20pt\hangafter=1}}

\gdef\refis#1{\item{#1.\ }}                     

\gdef\journal#1, #2, #3 {               
    {\it #1}, {\bf #2}, #3.}            




\def\endreferences{\body}

\def\figurecaptions             
  {\endpage \beginparmode \head{Figure Captions}
   \parskip=3pt \everypar{\hangindent=20pt\hangafter=1} }

\def\endpage                    
  {\vfill\eject}

\def\endpaper   {\endmode\vfill\supereject}
\def\endjnl     {\endpaper\end}


\def\ref#1{ref.{#1}}                    
\def\Ref#1{Ref.{#1}}                    
\def\[#1]{[\cite{#1}]}
\def\cite#1{{#1}}


\def\(#1){(\call{#1})}
\def\call#1{{#1}}
\def\frac#1#2{{#1 \over #2}}
\def\half{  {\frac 12}}

\def\fourth{{\frac 14}}
\def\12{{1\over2}}

\def\sla{\raise.15ex\hbox{$/$}\kern-.57em}
\def\leaderfill{\leaders\hbox to 1em{\hss.\hss}\hfill}
\def\twiddle{\lower.9ex\rlap{$\kern-.1em\scriptstyle\sim$}}
\def\bigtwiddle{\lower1.ex\rlap{$\sim$}}
\def\gtwid{\mathrel{\raise.3ex\hbox{$>$\kern-.75em\lower1ex\hbox{$\sim$}}}}
\def\ltwid{\mathrel{\raise.3ex\hbox{$<$\kern-.75em\lower1ex\hbox{$\sim$}}}}
\def\square{\kern1pt\vbox{\hrule height 1.2pt\hbox{\vrule width 1.2pt\hskip 3pt
   \vbox{\vskip 6pt}\hskip 3pt\vrule width 0.6pt}\hrule height 0.6pt}\kern1pt}
\def\tdot#1{\mathord{\mathop{#1}\limits^{\kern2pt\ldots}}}

\def\pmb#1{\setbox0=\hbox{#1}%
  \kern-.025em\copy0\kern-\wd0
  \kern  .05em\copy0\kern-\wd0
  \kern-.025em\raise.0433em\box0 }

\catcode`@=11
\newcount\r@fcount \r@fcount=0
\newcount\r@fcurr
\immediate\newwrite\reffile
\newif\ifr@ffile\r@ffilefalse
\def\w@rnwrite#1{\ifr@ffile\immediate\write\reffile{#1}\fi\message{#1}}

\def\writer@f#1>>{}
\def\referencefile{
  \r@ffiletrue\immediate\openout\reffile=\jobname.ref%
  \def\writer@f##1>>{\ifr@ffile\immediate\write\reffile%
    {\noexpand\refis{##1} = \csname r@fnum##1\endcsname = %
     \expandafter\expandafter\expandafter\strip@t\expandafter%
     \meaning\csname r@ftext\csname r@fnum##1\endcsname\endcsname}\fi}%
  \def\strip@t##1>>{}}

\def\citeall#1{\xdef#1##1{#1{\noexpand\cite{##1}}}}
\def\cite#1{\each@rg\citer@nge{#1}}     

\def\each@rg#1#2{{\let\thecsname=#1\expandafter\first@rg#2,\end,}}
\def\first@rg#1,{\thecsname{#1}\apply@rg}       
\def\apply@rg#1,{\ifx\end#1\let\next=\relax
\else,\thecsname{#1}\let\next=\apply@rg\fi\next}

\def\citer@nge#1{\citedor@nge#1-\end-}  
\def\citer@ngeat#1\end-{#1}
\def\citedor@nge#1-#2-{\ifx\end#2\r@featspace#1 
  \else\citel@@p{#1}{#2}\citer@ngeat\fi}        
\def\citel@@p#1#2{\ifnum#1>#2{\errmessage{Reference range #1-#2\space is bad.}
    \errhelp{If you cite a series of references by the notation M-N, then M and
    N must be integers, and N must be greater than or equal to M.}}\else%
 {\count0=#1\count1=#2\advance\count1
by1\relax\expandafter\r@fcite\the\count0,%
  \loop\advance\count0 by1\relax
    \ifnum\count0<\count1,\expandafter\r@fcite\the\count0,%
  \repeat}\fi}

\def\r@featspace#1#2 {\r@fcite#1#2,}    
\def\r@fcite#1,{\ifuncit@d{#1}          
    \expandafter\gdef\csname r@ftext\number\r@fcount\endcsname%
    {\message{Reference #1 to be supplied.}\writer@f#1>>#1 to be supplied.\par
     }\fi%
  \csname r@fnum#1\endcsname}

\def\ifuncit@d#1{\expandafter\ifx\csname r@fnum#1\endcsname\relax%
\global\advance\r@fcount by1%
\expandafter\xdef\csname r@fnum#1\endcsname{\number\r@fcount}}

\let\r@fis=\refis                       
\def\refis#1#2#3\par{\ifuncit@d{#1}
    \w@rnwrite{Reference #1=\number\r@fcount\space is not cited up to now.}\fi%
  \expandafter\gdef\csname r@ftext\csname r@fnum#1\endcsname\endcsname%
  {\writer@f#1>>#2#3\par}}

\def\r@ferr{\endreferences\errmessage{I was expecting to see
\noexpand\endreferences before now;  I have inserted it here.}}
\let\r@ferences=\references
\def\references{\r@ferences\def\endmode{\r@ferr\par\endgroup}}

\let\endr@ferences=\endreferences
\def\endreferences{\r@fcurr=0
  {\loop\ifnum\r@fcurr<\r@fcount
    \advance\r@fcurr by 1\relax\expandafter\r@fis\expandafter{\number\r@fcurr}%
    \csname r@ftext\number\r@fcurr\endcsname%
  \repeat}\gdef\r@ferr{}\endr@ferences}


\let\r@fend=\endpaper\gdef\endpaper{\ifr@ffile
\immediate\write16{Cross References written on []\jobname.REF.}\fi\r@fend}

\catcode`@=12

\citeall\refto          
\citeall\ref            %
\citeall\Ref            %


\def\fourth {{1\over 4}}
\doublespace
\vglue 0. truein
\title
{
Production of Vortices by Scattering Particles
}
\smallskip
\author
{Tanmay Vachaspati}
\affil
{
Tufts Institute of Cosmology, Department of Physics and Astronomy,
Tufts University, Medford, MA 02155.
}

\abstract
\doublespace

We give an action that can be used to describe the production of
global vortices in 2+1 dimensions by scattering
Nambu-Goldstone bosons. At strong self-coupling the action reduces to
scalar QED with particular values of the coupling constants and
the production cross-section is explicitly found. We also
consider the production of gauged vortices by scattering particles that have
an Aharanov-Bohm interaction with the vortex.

\endtopmatter

Solitons occur in a wide range of condensed matter systems and also
in many particle physics models. Usually a soliton is treated as
an object separate from the particle-like excitations that the theory
contains and the connection between particles and solitons remains
an unsolved problem. In particular, it is not known how to calculate
the cross-section for producing solitons in particle collisions.
In this paper, we shall attempt to establish what is at least a partial
connection by first considering the production of global vortices
in the scattering of Nambu-Goldstone bosons and then considering
the production of gauged vortices in the scattering of particles
having an Aharanov-Bohm interaction with the vortices. Our analysis
only applies to 2+1 dimensions.

The simplest model that gives rise to global vortices is described by
the action:
$$
S_{gl} = \int d^3 x \left [ \half |\partial_\mu \phi |^2 -
       {\lambda \over 4} ( |\phi |^2 - \eta ^2 )^2 \right ]
\eqno (1)
$$
where $\phi$ is a complex scalar field and the Greek indices run from
0 to 2. If we write
$$
\phi = \rho e^{i \alpha} \ ,
\eqno (2)
$$
the field $\alpha$ is massless and is the Nambu-Goldstone boson of the theory.
The field $\rho$ acquires a vacuum expectation value and the excitations
of $\rho$ about this value are the massive degrees of freedom with mass
$m_\rho = \sqrt{2 \lambda} \eta$.
(Note that, since we are working in 3 dimensions,
$\lambda$ has dimensions of mass and $\eta$ of $\sqrt{\rm mass}$.)
The equation of motion for $\alpha$ is
$$
\partial^\mu [ \rho ^2 \partial_\mu \alpha ] = 0
\eqno (3)
$$
and can be solved by setting
$$
\rho^2 \partial _\mu \alpha = \half \eta \epsilon_{\mu \nu \lambda}
                                       H^{\nu \lambda}
\eqno (4)
$$
where
$$
H^{\nu \lambda} = \partial ^\nu B^\lambda - \partial ^\lambda B^\nu
\eqno (5)
$$
for arbitrary gauge potentials $B^\mu$.

This connection between a massless scalar field and the dual tensor
field is well-known and has been used extensively to discuss the behaviour
of global vortices\refto{lr, ew, avtv, rdps}. We can write the action
$S_{gl}$ in terms of the gauge field $B_\mu$ as\refto{rdps}:
$$
S_{gl} = \int d^3 x \left [
      \half ( \partial _\mu \rho )^2 -
      {\lambda \over 4} (\rho^2 - \eta^2 ) ^2
          - \fourth {{\eta^2} \over {\rho^2}} H_{\mu \nu} H^{\mu \nu}
                    \right ] +
              g \int dx_v ^\mu B_\mu
\eqno (6)
$$
where, $x_v ^\mu$ is the position of the vortex and
the interaction between $B_\mu$ and the vortex is given
by the coupling constant\refto{ew, avtv}
$$
g = 2\pi \eta \ .
\eqno (7)
$$
In writing (6), no approximation has been made regarding the structure
of the vortex; the only step in going from (1) to (6) has been the
transformation of variables from $\alpha$ to $B_\mu$. In doing so,
the vortex degree of freedom had to be introduced so as to enforce the
condition that $\alpha$ must wind around the vortex. In (6) we have
assumed the presence of only one vortex; the presence of several
vortices can be accommodated by including a sum over vortices in the
interaction term. But we want to go further: we want
to include the possibility that the number of vortices in the system
can change. Hence, we must replace the vortex degree of freedom
with a vortex field $V$ that can be second quantized and whose
quanta would represent vortices. To do this,
note that the interaction of the vortex with $B_\mu$ is precisely
the interaction of a current, $g {\dot x}_v ^\mu$, with the gauge
field $B_\mu$. As things stand, the vortex is only identified with
the point around which $\alpha$ winds and hence the vortex
is massless. Therefore the vortex field is to be added
such that it gives the correct current-gauge
field interaction, is massless and respects the gauge invariance:
$$
B_\mu \rightarrow B_\mu + \partial_\mu \Lambda \ .
\eqno (8)
$$
(This gauge symmetry has nothing to do with
the global U(1) symmetry of $S_{gl}$ but is purely due to an introduction
of redundant variables in the transformation (4).)
This leads to the action:
$$
S_{gl+V} = \int d^3 x \left [
  \half ( \partial _\mu \rho )^2 -
   {\lambda \over 4} (\rho^2 - \eta^2 )^2 +
          \left | ( \partial_\mu +i g B_\mu )V \right |^2 -
    \fourth {{\eta^2} \over {\rho^2}} H_{\mu \nu} H^{\mu \nu}
               \right ] \ .
\eqno (9)
$$

In $S_{gl+V}$, a vortex only labels a point around which the original
field $\alpha$ circulates and arbitrary excitations of the structure of the
classical vortex are included since we have retained the field $\rho$.
The most unusual feature of $S_{gl+V}$ is the $\eta ^2 \rho^{-2}$
in front of the field strength term and, the physical interpretation
of this term is that $\eta^2 \rho^{-2}$ plays the role of a dynamical
dielectric constant and $\eta ^{-2} \rho ^2$ the role of a dynamical
magnetic permeability.

The action $S_{gl+V}$ is our central result. In $S_{gl+V}$, the vortices
appear explicitly as ordinary quanta would in a conventional quantum field
theory. By second quantizing the field $V$ it should be possible to
describe the creation and annihilation of vortices. However, the
peculiar interaction of $\rho$ with $B_\mu$ does not immediately lend
itself to a simple treatment of processes such as vortex production and
therefore it is necessary to consider various approximation schemes.

The simplest approximation to consider is the case when the vortex is
completely classical and the structure of the vortex can be considered
to be rigid. Then we may take $\rho$ to be always given by the classical
solution $\rho_v$ and integrate out the $\rho$ degree of freedom.
If we do this in (6), the vortex gets a mass, and the result is the
Kalb-Ramond action\refto{kr} (suitably reduced from 4 to 3 dimensions):
$$
S_{KR} = - \mu \int d\tau + g \int dx_v ^\mu B_\mu -
                      \fourth \int d^3 x H_{\mu \nu} H^{\mu \nu}
\eqno (10)
$$
where $\tau$ is the proper time of the vortex. The energy per unit length
of the vortex is\refto{rdps}
$$
\mu = \int d^2 x \left [ \half (\partial _i \rho_v )^2 -
     \half \rho_v ^2 \left ( 1 - {{\rho_v}^2 \over \eta^2} \right )
             (\partial _i \alpha_v )^2 +
         {\lambda \over 4} (\rho_v ^2 - \eta^2 )^2 \right ] \
\eqno (11)
$$
The fields labelled by a subscript $v$ refer to the field
configurations for a static vortex and the index $i = 1,2$. By numerically
solving for the classical vortex profile, we have found $\mu$:
$$
\mu = 0.18 \pi \eta^2 \sim \eta^2
\eqno (12)
$$
and it is easy to see, by rescaling arguments, that $\mu$ is
independent of $\lambda$.

Now we introduce the vortex field in the Kalb-Ramond action. This simply
leads to massive scalar QED without any vortex self-interactions:
$$
S_V = \int d^3 x \left [
          \left | ( \partial_\mu +i g B_\mu )V \right |^2 -
   \mu^2 |V|^2 - \fourth H_{\mu \nu} H^{\mu \nu}
               \right ] \ .
\eqno (13)
$$

With this action, it is possible to consider the production of vortices -
quanta of $V$ - in processes involving Nambu-Goldstone bosons - that is,
quanta of $B_\mu$ (denoted by $\gamma$). The calculations that need to be
done are standard scalar QED calculations. At high energies, perturbative
calculations can be done and the lowest order cross-section for
$\gamma \gamma \rightarrow VV$ is:
$$
{{d\sigma} \over {d\theta}} = \left ( {{g^2} \over {E}} \right ) ^2
    {1 \over {4\pi E}} \left [ 1 -
{{2 (1-s^2) sin^2 \theta} \over
 {(1-s^2 ) sin^2 \theta + s^2 } }
                       \right ] ^2
\eqno (14)
$$
where, $E \ge 2\mu $ is the energy in the center of mass system,
$s = 2\mu /E \le 1$,
$\theta$ is the angle between the incoming and outgoing beams and the
$\gamma$ polarizations have been taken to be aligned.

The production of vortices requires an energy $E > 2\mu$ while,
in deriving the Kalb-Ramond action, we have used $E < m_\rho$ since we
have assumed that the vortex structure is rigid and, hence,
quanta of the field $\rho$ in $S_{gl}$ remain unexcited.
Therefore our result for the production cross-section is only valid
in the regime where $m_\rho > 2 \mu$,
or, in terms of the parameters, where $\sqrt{\lambda} \gtwid \eta$.
This is in the strong coupling regime of the theory described by $S_{gl}$.
The only way out of the strong coupling regime would seem to be
to stick with the action $S_{gl+V}$ or derive some approximation scheme
that interpolates between the two extreme cases represented by
$S_{gl+V}$ and $S_{V}$.

The reader might wonder if the
result (14) is believable since generally not much can be said of theories
at strong coupling. However, in our case, the only step in going from (9)
to (13) is an integration over the massive degree of freedom $\rho$ and,
in particular, no perturbative expansion in $\lambda$ has been utilised.
Also note that a simple rescaling argument is sufficient to show that $\mu$
is independent of $\lambda$ and so the mass of the vortex does not depend
on the value of the coupling constant. Once we have obtained (13), the
result (14) follows within the energy domain $m_\rho > E > 2 \mu$. It is
also helpful to repeat this argument in terms of quantum fluctuations about
the vortex. The fluctuations will consist of zero modes corresponding to
translations of the vortex and of massive modes corrresponding to fluctuations
of the vortex profile. The latter fluctuations can be viewed as massive
$\rho$ quanta in the background of the vortex. In the strong coupling
regime, the mass of $\rho$ becomes large and hence the massive fluctuations
cannot be excited at the energy scales we are interested in. Therefore
the zero modes are the the only remaining fluctuations and these are already
included in (10) as the dynamical degrees of freedom of the vortex.

An observation relevant to this
discussion is that some features of $~{}^4He$ can be described using
the Kalb-Ramond action even though ${}^4He$ is in the strong coupling
regime with (see Ref. \cite{rd}) $\lambda_4 \sim 10^9$. (The subscript 4 on
$\lambda$ indicates that the coupling is for the theory in 4 dimensions.)
Hence, it may be possible to test the validity of this approach by doing
experiments in the laboratory.

Observe that there is no {\it classical} production of vortices in the model
(13) and also that, since we are working in three dimensions, the theory is
confining. The confinement is most simply seen by noting that the
electric force between two charges in 3 dimensions falls of as $1/r$ and the
potential diverges logarithmically. A consequence of the weak confining force
is that, after the vortices are produced, they will not move away indefinitely
from each other but will eventually turn around and recollapse. However,
we may define that vortices are produced whenever the separation of the
vortices exceeds the width of the vortices (roughly equal to $m_\rho ^{-1}$).
Then the logarithmic inter-vortex potential has an effect on the dynamics
of the vortices that have already been produced but does not affect
the production process itself.

Outside the strong coupling regime,
a method that suggests itself is to treat the vortex as consisting of
an inner core that remains classical and an outer shell which is
quantum. Suppose that the classical core has a radius $R$. Inside
the classical core we can take $\rho = \rho_v$ as was done in the
Kalb-Ramond case
while outside the core we write
$$
\rho = \eta + \sigma \ .
\eqno (15)
$$
Then, the full action $S_{gl+V}$ may be approximated by,
$$
\eqalign{
S_{V+\sigma } = \int d^3 x \biggl [
 | (\partial_\mu + i g B_\mu )V |^2 - \mu_R ^2 |V|^2 &
 -\fourth \left ( 1 + {\sigma \over \eta} \right )^{-2}
                                          H_{\mu \nu} H^{\mu \nu}
\cr
&
 + \half (\partial_\mu \sigma )^2 -
                {\lambda \over 4} ( 2 \eta + \sigma ) ^2 \sigma^2
                                  \biggr ]
}
\eqno (16)
$$
where, $\mu_R$ is the mass of the classical vortex within the core of
radius $R$ and we have ignored terms of order $E^2 R^2$ and higher
where $E$ is the energy scale of interest.
The $\sigma$ field represents the dressing on the
bare classical vortex and we can expand the coefficient of the field
strength term in increasing powers of $\sigma /eta$. (We expect that this
expansion will make sense since $|\sigma | / \eta < 1$.) An extreme
case of (16) is when $R=0$ and the whole vortex is considered as
being dressing. In this case, $\mu_R =0$ and, at the center of the
vortex, we necessarily have $\sigma = -\eta$. Therefore, if we
expand out the coefficient of the field strength term, we would be
forced to consider the entire expansion with its infinite number of
terms. The general policy is that we would like $R$ to be small
compared to $E^{-1}$ but not too small, as that would entail retaining
large orders in $\sigma /\eta$. Another extreme case is the large coupling
case already considered where we can simultaneously take $E^2 R^2$
to be small and also truncate the $\sigma$ expansion to the lowest order.

A physical picture of the production process is that the $\gamma \gamma$
collision produces the innermost cores (of size $R$ such that $ER < 1$)
of the vortex-anti-vortex pair. The relative momenta with which the
vortices in the pair are moving away from each other
at the moment of production is $\sqrt{E^2 - 4 \mu_R ^2}$. As the
pair is trying to separate, the $B_\mu$ and $\sigma$ fields provide
an attractive potential that tries to prevent the separation.
This attraction is due to the need to create the outer shell
of the vortices and the accompanying gauge field configuration
as the vortices move apart. If
the initial momenta of the innermost cores is large enough to overcome
the attractive potential and the energy of the inner cores is not
transferred to excitations of the outer shell of the vortices
or quanta that can be radiated away,
the pair will separate out to distances larger than the vortex
core width and we can say that a vortex pair has been produced.
We know that the total energy required to separate the pair beyond
the core width is $2(\mu - \mu_R )$ and so we at least need
$E > 2 \mu$ to create the vortices. However the
rate at which the energy of the vortices is lost to excitations
of the outer shells or to quanta that are radiated away
will depend on the coupling $\lambda$ and needs to be calculated
from action such as $S_{V+\sigma}$. The only case
when quanta of $\sigma$ will not be excited is when
the residual energy after producing the inner cores is not large
enough to produce any $\sigma$ quanta: $E /2 - \mu_R < m$.
At threshold, $E/2 = \mu$ and
then, a translation of the inequality in terms of the coupling
constants leads to: $\sqrt{\lambda} /\eta > 1 - \mu_R /\mu$.
If we take $R = \eta^{-2}$, and evaluate $\mu_R /\mu$ numerically,
we find that the inequality is satisfied for
$\sqrt{\lambda} /\eta > 0.7$.

At first sight, the result (14) seems puzzling as one expects that the
vortex production cross-section should be exponentially suppressed since
the vortex is an extended object. However, there is no exponential
suppression in (14) since the result is only valid in the thin vortex
limit where the vortex can be treated as if it had no structure.
To find the production cross-section in the case when the vortex does
have structure, one would have to find
methods to analyze the action $S_{gl+V}$ for arbitrary values of $\lambda$.
What has emerged from the analysis here is that the production cross-section
must go over to (14) in the limit that $\lambda$ becomes large.


Next we shall study the possibility of producing gauged vortices.
Consider the Abelian-Higgs model:
$$
S_{AH} = \int d^3 x \left [
          \left | ( \partial_\mu + i e A_\mu )\phi \right |^2 -
    \fourth F_{\mu \nu} F^{\mu \nu} - \lambda (|\phi |^2 - \eta^2 )^2
               \right ] \ .
\eqno (17)
$$
It is well-known that this model admits vortex solutions\refto{hnpo}. But
now, when $\phi$ gets an expectation value, all the fields become
massive. In the limit that the masses are very large compared to the
energies we are interested in, the action for the vortices is simply
that of a point particle:
$$
S_p = - \mu \int d\tau
\eqno (18)
$$
where, as before, $\mu \sim \eta^2 $ is the vortex mass. ($\mu$ also
depends weakly on the ratio $e^2 /\lambda$ which we assume
to be of order one.) Therefore
if we replace the single vortex by a field of vortices and ignore
vortex-vortex interactions, the result is a free field theory:
$$
S_{free} = \int d^3 x [ (\partial_\mu V )^2 - \mu^2 |V|^2 ] \ ,
\eqno (19)
$$
and vortices cannot be produced in this model because there are no
particles to produce them with. Therefore to get something interesting,
it is necessary to introduce other fields in the Abelian-Higgs model
or to consider excitations of the scalar and vector fields already
present in $S_{AH}$. Here we will only consider the first of these
two options.

It should be mentioned that one can always introduce gravity in the
model (19) and hope to get vortex production in a non-trivial gravitational
background. In four spacetime dimensions this has already been considered
in the context of nucleation of topological defects in de Sitter
space\refto{rbagav}.

Let us consider the case when there is an extra scalar field present
in $S_{AH}$:
$$
S_{AH+} = S_{AH} + \int d^3 x \left [
       \half \left | ( \partial_\mu + i e' A_\mu )\psi \right |^2 -
                U(|\phi |, |\psi |)   \right ] \ .
\eqno (20)
$$
It is important to note that the gauge coupling $e'$ is chosen to be
different from $e$. Also, the modulus of the scalar field $|\phi |$
occurring in the potential may be replaced by the vacuum expectation value
$\eta$ in the zero width vortex limit.

Now let us integrate out the massive degrees of freedom:
$$
S_{AH+} \rightarrow - \mu \int d\tau + \int d^3 x \left [
 \half \xi^2 \left \{ \partial_\mu \left ( \alpha - {{e'} \over e} \theta
                               \right ) \right \} ^2 +
   \half (\partial_\mu \xi )^2 - U(\xi ) \right ] \ ,
\eqno (21)
$$
where, we have used (18), together with
$$
A_\mu = - {1 \over e} \partial_\mu \theta  \ ,
\eqno (22)
$$
where, $\theta$ is the polar angle about the vortex, and
$$
\psi = \xi e^{i \alpha } \ .
\eqno (23)
$$

We repeat the same steps that led to the Kalb-Ramond action for the
global vortex and define:
$$
\xi^2 \partial _\mu \left ( \alpha - {{e'} \over e} \theta \right ) =
            \half \eta ' \epsilon_{\mu \nu \lambda} H^{\nu \lambda}
\eqno (24)
$$
where $(\eta ') ^2$ is an arbitrary mass scale. This transformation
of variables leads to a Kalb-Ramond type of action:
$$
S_{KR+} = - \mu \int d\tau + q \int dx^\mu B_\mu +
  \int d^3 x \left [ -\fourth {{{\eta '}^2} \over {\xi ^2}}
                                       H_{\mu \nu} H^{\mu \nu} +
   \half (\partial_\mu \xi )^2 - U(\xi ) \right ] \ ,
\eqno (25)
$$
where, the charge $q$ will now be determined.

If we wish to find a non-trivial field configuration $\psi$ in the
background of the vortex, single-valuedness of $\psi$ requires that
$$
\oint_{C_\infty} dx^\mu \partial_\mu \alpha = 2\pi n
\eqno (26)
$$
for some integer $n$. Therefore, if we let
$$
\beta = \alpha - {{e'} \over e} \theta
\eqno (27)
$$
then, a constraint on the field configuration $\psi$ is
$$
\oint_{C_\infty} dx^\mu \partial_\mu \beta =
                2\pi \left ( n - {{e'} \over e} \right )
\eqno (28)
$$
where, $C_\infty$ denotes a closed contour encircling the vortex.
We will only consider the case when $e' /e$ is not an integer and
hence, the quanta of $\psi$ have an Aharanov-Bohm interaction with the
vortex.

The constraint (28) can now be expressed in terms of the gauge fields
introduced in (24) by the manipulations in Refs. \cite{ew, avtv}.
This results in the coupling of the gauge field $B_\mu$ with the vortex
as in (25) and where the charge $q$ is given by
$$
q = 2\pi \left [ n -{{e'} \over e} \right ] \eta ' \ .
\eqno (29)
$$
Furthermore, we expect the most weakly coupled field configurations
to be the most important and so $n$ should be chosen to minimize $q$.

Next we wish to introduce the vortex field $V$ so that we can describe
vortex production. This is done using the minimal prescription and
ignoring vortex-vortex interactions:
$$
S_{V+} = \int d^3 x \left [
          \left | ( \partial_\mu +i q B_\mu )V \right |^2 - \mu^2 |V|^2 -
    \fourth {{{\eta '}^2} \over {\xi ^2}} H_{\mu \nu} H^{\mu \nu} +
   \half (\partial_\mu \xi )^2 - U(\xi ) \right ] \ .
\eqno (30)
$$
Here ${\eta '} ^2 \xi^{-2}$ plays
the role of a dynamical dielectric constant and
${\eta '}^{-2} \xi ^2$ the role of
magnetic permeability.
In contrast to the global case (see below eq. (9)),
$\xi =0$ in vacuum, and so the dielectric constant
is infinite while the magnetic permeability is zero. Consequently the
electric and magnetic fields must vanish in the vacuum.

The situation becomes more interesting if we consider the case when
the potential $U(\xi )$ can be ignored, and a background of the field
$\psi$ is present:
$$
\psi = A e^{i\alpha (t, \vec x ) }
\eqno (31)
$$
where, $A$ is a constant. (This could happen if we were to consider the
scattering of waves of $\alpha$, or, if there were a non-trivial
potential $U(\xi )$ and the global
U(1) symmetry associated with phase rotations of $\psi$ was broken,
giving a vacuum expectation value to $\psi$.)
Then we have $\xi = A$ and the relevant action (30) is scalar QED once
again. Hence the cross-section for the production of vortices is
identical to that in the global vortex case except for the value of the
charge. The cross-section will be given by (14) with
$ g = - 2\pi A e' /e$, where we have taken $e'$ to be much smaller
than $e$ and set $n=0$.

The model (30) can be useful for calculating the production rates of
vortices only in the strong coupling regime (large $e$ and $\lambda$).
If these couplings are not large, however,
we would have to retain the possibility that quanta of the massive
fields $\phi$ and $A_\mu$ could be produced in the scattering of
$\psi$ particles. For this, we would have to deal with an action
similar to $S_{gl+V}$ in (9). This action can be written down by
adding $S_{AH}$ to $S_{V+}$, including the possibility of
$\rho$, $\xi$ interactions in the potential and setting $\mu = 0$:
$$
S_{AH+V} = S_{AH} + \int d^3 x \left [
          \left | ( \partial_\mu +i q B_\mu )V \right |^2  -
    \fourth {{{\eta '}^2} \over {\xi ^2}} H_{\mu \nu} H^{\mu \nu} +
   \half (\partial_\mu \xi )^2 - U(\rho , \xi ) \right ] \ .
\eqno (32)
$$

How can one extend this analysis to 3+1 dimensions?
Clearly the 2+1 dimensional results can describe vortices that are
infinite along the added third spatial dimension. In a realistic
set-up, however, the scattering is done with finite beam sizes and the
vortices cannot be infinite. The vortices cannot end and hence would have
to form a closed loop. To deal with this situation, one would have to
construct a formalism capable of describing the creation and annihilation
of one-dimensional extended systems.
In the context of vortices in condensed matter systems or fluids, the system
is finite and it may be possible to create vortices that span the whole
system. Such vortices are effectively infinite.

The ideas described above rest on the duality transformation relating
a scalar to a gauge field. While the transformation may be reliable at
a classical level, it is not certain that the transformation will give
equivalent theories even at the quantum level. On the other
hand, however, these ideas may be testable in the laboratory if (1),
or some refinement of it, provides an adequate description of fluids or
condensed matter systems such as ${}^4He$. This remains to be seen. Another
possibility that comes to mind is the production of electroweak
vortices in the scattering of leptons\refto{tv}. All the ingredients
for the production of gauged vortices seem to be present in the electroweak
model: vortex solutions and particles interacting with the vortex via
an Aharanov-Bohm effect are also present. There are some complications
too: the vortices are not topological and the particles are fermions.
Another requirement that is not met, is the strong coupling limit and
methods will have to be found to be able to treat actions such as
$S_{gl+V}$ and $S_{AH+V}$. Efforts in this direction are in progress.

\

\

\noindent {\it{Acknowledgements:}}

I am grateful to Sidney Coleman, Tom Kibble, Samir Mathur, Poul Olesen,
John Preskill, Qaisar Shafi and Alex Vilenkin for comments. This work
was supported by the NSF.

\vfill

\references

\refis{lr} F. Lund and T. Regge, Phys. Rev. D14, 1524 (1976).

\refis{kr} M. Kalb and P. Ramond, Phys. Rev. D9, 2273 (1974).

\refis{ew} E. Witten, Phys. Lett. B158, 243 (1985).

\refis{avtv} A. Vilenkin and T. Vachaspati, Phys. Rev. D34, 1138 (1985).

\refis{rdps} R. L. Davis and E. P. S. Shellard, Phys. Lett. B214,
219 (1988).

\refis{rd} R. L. Davis, Physica B178, 76 (1992).

\refis{hnpo} H. B. Nielsen and P. Olesen, Nucl. Phys. B61, 45 (1973).

\refis{rbagav} R. Basu, A. H. Guth and A. Vilenkin, Phys. Rev. D44,
340 (1991).

\refis{tv} T. Vachaspati, Nucl. Phys. B397, 648 (1993).

\endreferences

\vfill
\eject


\endjnl
\end